\documentclass[iop]{emulateapj} 
\shortauthors{Sekanina}
\shorttitle{Prognostication of Bright Kreutz Sungrazers}
\slugcomment{Version \today }

\begin{document}
\title{On the Problem of Prognostication of Bright Kreutz Sungrazers}
\author{Zdenek Sekanina}
\affil{La Canada Flintridge, California 91011, U.S.A.; {\sl ZdenSek@gmail.com}}

\begin{abstract} 
Tidal fragmentation at perihelion and nontidal fragmentation elsewhere
cause the orbital distribution of Kreutz sungrazers of all sizes to be
extremely complicated and highly nonuniform.  Among the features are
(largely fortuitous) clusters of bright (naked-eye) objects and clumps
of dwarf objects (often closely genetically related, as their detection
primarily by the SOHO coronagraphs suggests) on the one hand; and both
spectacular and less brilliant sibling sungrazers, whose perihelion
times are scattered over centuries, on the other hand.  Investigation
of four fragment nuclei of the Great September Comet of 1882, the
products of a perihelion breakup of the comet's original nucleus,
showed that their orbital periods followed a distinct pattern, which
likewise applied to other tidally split sungrazers and was characterized
by a specific value of the second difference of parameter~$u_{\rm frg}$
of neighboring fragments' centers of mass.  The algorithm has a potential
for the prognostication of bright Kreutz sungrazers over the rest of the
21st century and beyond.  However, because of its as yet unverified
empirical character, the utmost caution should be exercised when applying
the procedure.{\vspace{-0.05cm}}
\end{abstract}
\keywords{individual comets:\,372\,BC,\,252,\,363,\,467,\,852,\,1041,\,X/1106\,C1,\,1138,\,C/1843\,D1,\,C/1880~C1, C/1882\,R1,\,C/1963\,R1,\,C/1965\,S1,\,C/1970\,K1,\,C/2011\,W3,\,others; methods:\ data analysis\vspace{-0.2cm}}
\section{Introduction} 
\vspace{-0.03cm}
When a spectacular comet (known nowadays by its designation C/1843~D1)
appeared in close proximity of the Sun in late February and early March
1843, a popular view among comet astronomers of the time was that
this was the returning comet of 1668 (in today's catalogues designated
C/1668~E1), implying an orbital period of 175~years.  When another
brilliant comet (now known as C/1880~C1) in essentially the same
orbit appeared very close to the Sun at the beginning of February
1880, a widely accepted view was that this object was the returning
sungrazer of 1843, the orbital period having shortened, somewhat
strangely, to 37~years.  But when a phenomenal sungrazer (now known
as C/1882~R1) appeared in September 1882, moving again in a very
similar orbit, an orbital period of 2.64~years was fortunately not
invoked.  Instead, numerous orbit determinations of this rather
extensively observed object confirmed that its period was in fact
on the order of several hundred years (e.g., Fabritius 1883, Frisby
1883, Morrison 1883).  The implication was that the sungrazers were
independent objects, fragments of a common parent rather than returns of
the same comet.  The arrival, in January 1887, of a headless sungrazer
(designated C/1887~B1) meant that three naked-eye objects in similar
orbits greatly approaching the Sun were observed over a period of 7~years
and four over a period of 44~years.  This crop followed a period of
140~years, in the course of which no such comets displaying long tails
were seen at all.

Nearly 59 years after the comet of 1887, a sungrazer (du Toit, C/1945~X1)
was discovered and observed telescopically over a period of several days.
Finally, some 18~years later, another naked-eye sungrazer appeared in
September 1963 (known as comet Pereyra, C/1963~R1), followed by the most
spectacular comet of the 20th century (Ikeya-Seki, C/1965~S1) in the
months starting with September of 1965; and by another sungrazer
(White-Ortiz-Bolelli, C/1970~K1) seen with the naked eye briefly in
May 1970.  Thus, three naked-eye objects in sungrazing orbits
appeared once again over a period of 7~years and four (one telescopic)
over a period of 25~years.

To maintain that the distribution of perihelion times of these sungrazers,
named after the German astronomer Heinrich Kreutz, was uneven is a gross
understatement.  Yet, no alarm had been expressed about the enormous
fluctuations in the rate of their appearances until attention to this
anomaly was called by Marsden (1967).

\section{Clusters of Bright Kreutz Sungrazers} 
Marsden pointed out that the ``{\it times of the appearances of the
comets are distributed in a highly nonuniform manner.  There seem to be
three distinct clusters~--- one in the late 17th century, a second in the
19th century, and a third in progress at the present time.}\footnote{That
is, in the 1960s.}''

A complication was that the term {\it cluster\/} was not defined.
Conservatively, the 19th century cluster may be understood to
have consisted of the three sungrazers between 1880 and
1887.\footnote{Another, much fainter sungrazer --- X/1882~K1 (Tewfik)
--- was seen with the naked eye during a total solar eclipse
on~1882~May~17.}  At the time Marsden's paper was published, the 20th century
cluster was made up by only two sungrazers, and even though he correctly
anticipated that a third was on the way, the temporal extent of this cluster
before 1970 was unclear.  The late 17th century cluster, stretching over
more than 12~years, was altogether an enigma.  According to Marsden it
consisted of C/1689~X1, C/1695~U1, and X/1702~D1, but Kreutz (1901) was
skeptical about the first two being sungrazers at all, with C/1689~X1
not apparently moving in a retrograde orbit (Holetschek 1892).  Only
comet X/1702~D1 was a likely sungrazer closely associated with the
spectacular comet of 1882.  Several candidates appeared in the 1660s,
including C/1668~E1 (Section~1), but this would be another cluster.
When linked with X/1702~D1, these objects would make up a cluster
extending over more than 30~years.  But the 20th century cluster should
then include C/1945~X1.  A slightly more relaxed definition of a
cluster would imply the existence of eight of them between $\sim$1550
and 1970 (Sekanina \& Chodas 2007), or one per 60~years on the average.
It was this compelling {\it empirical\/} evidence --- rather than any
theoretical construct --- that led us to predict, in that same paper,
that there were signs of ``{\it another cluster of bright Kreutz
\ldots comets \ldots on its way to the Sun in the coming decades, with
the earliest objects expected to arrive perhaps as soon as several
years from now\/}.\footnote{That is, several years from 2007.}''
Four years after this bold prognosis was made, comet Lovejoy
(C/2011~W3) appeared.  Whether it was or was not a member of the
predicted 21st century cluster depends again on the definition
of a cluster.  If a cluster is not to extend over a period longer
than 10~years, for example, then Lovejoy was not part of any cluster,
as no additional bright sungrazers appeared by 2021.  If a cluster may
stretch over as much as 30~years, we will be able to answer the question
of comet Lovejoy's cluster membership only in 2041.

\section{Tidal and Nontidal Fragmentation of Kreutz Sungrazers} 
Another issue that the existence of the clusters brought up was their
nature:\ were they integral part of the Kreutz group's architecture,
such as a corollary of the fragmentation process, or did they represent
merely products of coincidence?  Marsden (1967), who accepted Kreutz's
premise of the sungrazing comets fragmenting exclusively at perihelion,
concluded that their clusters were ``{\it largely fortuitous\/},''
because the process of tidal fragmentation offered no obvious mechanism
for generating clusters.

Marsden's (1989) subsequent orbital computations for fainter Kreutz
sungrazers, imaged by a coronagraph on board the {\it Solar Maximum
Mission\/} (SMM) in the 1980s, showed a tendency toward clumping
on a timescale of $\sim$10~days, which could not be satisfactorily
explained in terms of tidal fragmentation at close proximity to the
Sun.  The episodic character of the temporal distribution of {\it
dwarf\/} Kreutz sungrazers, imaged since 1996 by the coronagraphs
on board the {\it Solar and Heliospheric Observatory\/} (SOHO), has
demonstrated the existence of a complex sequence of recurring {\it
nontidal-fragmentation\/} events, characterized by low separation
velocities and occurring along the entire length of the orbit
(Sekanina 2000).  As the final products of the process of {\it
cascading fragmentation\/}, the dwarf sungrazers have often been
arriving in pairs or larger groups, separated from each other in
extreme cases by as little as a fraction~of~an~hour.

The concept of cascading fragmentation profoundly impacted the
direction of the research on Kreutz sungrazers.  The change that
the motion of any fragment was subjected to depended strongly on
the position in the orbit where the separation occurred.  Whereas
the fragment's only orbital element affected significantly during
a breakup at perihelion was its {\it orbital period\/}, the motion
of a fragment separating at aphelion (or, more generally, very far
from the Sun) was affected significantly in the {\it perihelion
distance\/} and {\it angular elements\/}, the {\it orbital period\/}
being modified only marginally.  This meant that the produced {\it
pair\/} or {\it cluster of fragments\/} was the tighter the closer
to aphelion the breakup episode took place.

To illustrate the differences between effects of tidal and nontidal
fragmentation on the orbital period, it is necessary to distinguish
between Marsden's Subgroups~I and II.  A massive fragment breaking off
{\vspace{-0.01cm}}tidally at perihelion with a separation velocity of
1~m/s in the direction of motion ends up in an orbit whose period is
145~years longer than its parent's if Subgroup~I, but only 119~years
longer than its parent's if Subgroup~II.  By contrast, fragments
separating from their parent of {\vspace{-0.01cm}}Subgroup~I with a
{\it radial\/} velocity of 1~m/s must be released at heliocentric
distances of at least 0.10~AU in order to form a cluster extending
over a period of 30~years at their next arrival at perihelion, but
greater than 1.8~AU in order to comprise a cluster that is to extend
over 7~years.  The numbers are very similar for{\vspace{-0.01cm}}
Subgroup~II.  If the fragments have a {\it transverse\/} separation
velocity of 1~m/s and no radial velocity, their release must take
place still closer to the Sun, at distances not smaller than 0.023~AU
for the cluster to extend over 30~years and at least 0.10~AU to
extend over 7~years in the case of Subgroup~I, and, respectively,
0.028~AU and 0.12~AU in the case of Subgroup~II.  The radial
component of the separation velocity is thus much more effective
in setting the fragments apart than the transverse component.  The
outgassing-driven nongravitational acceleration on the fragments'
motions is in all these scenarios assumed to be trivial.  If not, the
heliocentric distances at fragmentation must be still greater, the
fragments separating at later times.

The {\it clusters\/} of genetically related bright Kreutz sungrazers
extending over many years were thus born nontidally not too long after
perihelion.  By contrast, the times and heliocentric distances of
fragmentation events that gave birth to {\it clumps\/} of dwarf
Kreutz sungrazers,\footnote{I deliberately use different terms
--- {\it cluster\/} and {\it clump\/} --- to emphasize the
distinction between the bright and dwarf sungrazers.} detected
by the coronagraphs on board the SMM and SOHO spacecraft, were
at the same separation velocities very different.
{\vspace{-0.01cm}}Two such objects separating from each other with
a {\it radial\/} velocity of 1~m/s must have been released at a
heliocentric distance of $\sim$119~AU {\it after aphelion\/} in
order to arrive at perihelion 10~days apart, regardless of whether
they were members of Subgroup~I or II.  If the objects had a zero
{\vspace{-0.01cm}}radial velocity and a {\it transverse\/} velocity
of 1~m/s, they must have been released already at a heliocentric
distance of 25~AU {\it on their way to aphelion\/} in order to
arrive at perihelion 10~days apart, if they belonged to Subgroup~I,
but at 30~AU before aphelion, if they were members of Subgroup~II.
In order for two dwarf sungrazers of either subgroup to arrive at
perihelion only 1~day apart, they must have separated at 46~AU
from the Sun after aphelion,{\vspace{-0.01cm}} if their radial
velocity was 1~m/s, but very close to aphelion{\vspace{-0.01cm}}
if their transverse velocity was 1~m/s and radial velocity was zero.

\section{Types of Sungrazers' Clusters and Clumps} 
\vspace{-0.05cm}
A tacit condition for a cluster of bright sungrazers or a clump of dwarf
sungrazers formed genetically is that {\it all\/} its members belong to
the {\it same subgroup\/} --- this is a {\small \bf genuine cluster/clump}.
%
If at least two members belong to the same subgroup, but others do not,
I call the collection a {\small \bf mixed cluster/clump}.  If each
member is from a different subgroup, it is a {\small \bf fortuitous
cluster/clump}.\footnote{This of course can happen only when the number
of objects in a cluster/clump does not exceed the number of subgroups.}

Defining Marsden's\,(1967,\,1989)\,clusters as sets~of~three objects
each, one finds that the {\small \bf 19th century cluster} was {\small
\bf mixed}, as C/1880~C1 and C/1887~B1 belonged to Subgroup~I, while
C/1882~R1 to Subgroup~II; whereas the {\small \bf 20th century cluster}
was {\small \bf fortuitous}, as C/1963~R1 was Subgroup~I, C/1965~S1 was
Subgroup~II, and C/1970~K1 was Subgroup~IIa.  This conclusion essentially
parallels Marsden's.  The fortuitous nature of the pair of comets
Pereyra and Ikeya-Seki is in fact documented directly, because their
past temporal relationship is known rather accurately.  Ikeya-Seki
was at perihelion in 1138 (Seka\-nina \& Kracht 2022), while Pereyra's
barycentric orbital period was \mbox{$904 \!\pm\! 18$}~years (Marsden et
al.\ 1978), so that this comet was at perihelion nearly a century earlier.

Inspection of Marsden's catalogue of parabolic-orbit approximations
for more than 1500 dwarf Kreutz sungrazers, detected by the C2 and C3
coronagraphs on board the SOHO spacecraft between 1996 and mid-2010,
revealed a number of clumps extending over periods of up to several
days.  Examples are presented in Table 1, adapted from Sekanina
(2024a).  In sharp contradiction to the 19th and 20th century clusters
of bright sungrazers, all entries in the table are members of
Population~I (identical with Marsden's Subgroup~I), so that each of
the tabulated {\small \bf clumps} is {\small \bf genuine}.  Marsden's
orbital elements, referred to as {\it nominal\/}, were corrected to
fit the standard position of the line of apsides.  The corrected
elements are referred to as {\it true\/}.  The individual columns
provide:\ the object's designation (followed by an asterisk when
the object's orbit is not based exclusively on the astrometry from
{\vspace{-0.05cm}}C2 coronagraphic images); the true perihelion
time, $\widehat{t}_\pi$, that in general differs slightly from the
perihelion time derived by Marsden (see{\vspace{-0.08cm}} Sekanina
2024a for details); the true longitude of the ascending node,
$\widehat{\Omega}$; the {\vspace{-0.08cm}}difference between the nominal
and true nodal longitudes, \mbox{$\Omega \!-\! \widehat{\Omega}$}; the
nominal perihelion distance, $q$, in units of the Sun's radius, $R_\odot$;
and the time{\vspace{-0.07cm}} of the last image~in C2 relative to
perihelion, \mbox{$t_{\rm last} \!-\! \widehat{t}_\pi$}, in days.

Another reason why the tabulated clumps of dwarf Kreutz sungrazers cannot
be fortuitous, is that application of the Poisson distribution shows
tight pairs occurring much more often than is statistically expected.
For example, the pairs of C/2006~K13 and C/2006~K14 from Clump~H, and
C/2007~U3 and C/2007~U5 from Clump~B reach perihelion (a fraction of a day
after their disintegration) simultaneously, while C/2006~J7 and C/2006~J8
from Clump~E, and C/2007~L6 and C/2007~L7 from Clump~A arriving at perihelion
(also after disintegration) only 15~minutes apart.  It is possible that
Clumps~A, B, and G are closely related in that a first-generation
fragment broke up into two second-generation fragments, one of which
subsequently fell apart to create the third-generation fragments that make
up Clump~A, while the other split into two, each of which almost immediately
crumbled into the fragments of Clumps~B and G, respectively.  Similarly
related could be Clumps~D, E, and H, and also Clumps~F and J.

\begin{table} 
\vspace{0.15cm}
\hspace{-0.25cm}
\centerline{
\scalebox{0.999}{
\includegraphics{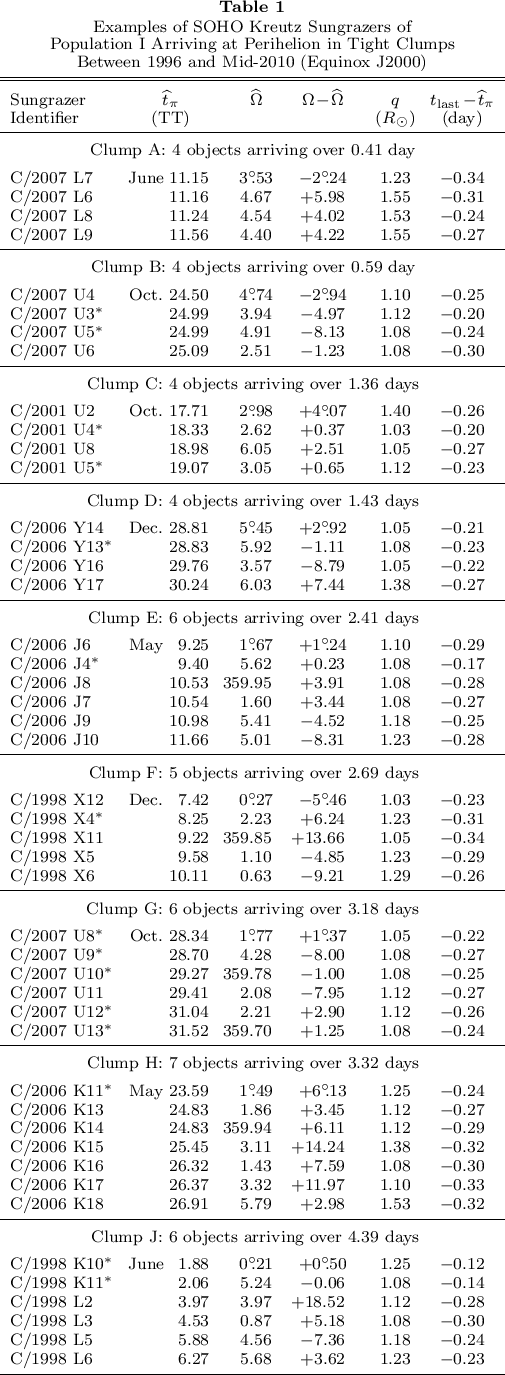}}}
\end{table}

One certainly can find rather tight fortuitous pairs and clumps of
dwarf sungrazers among the 1500+ entries in Marsden's catalogue, but
thanks to cascading fragmentation the genuine clumps dominate ---
the fundamental difference between the clusters of bright sungrazers
and the clumps of dwarf sungrazers.

\section{Tidal Fragmentation With No Momentum Exchange} 
In Section 3 I noted that the products of tidal fragmentation of a
sungrazer very near its perihelion end~up in paths of very different
orbital periods because of~low, meter-per-second velocities, at which
they separate~from the parent comet.  The greatest effect is triggered
by an impulse exerted in the direction of the orbital-velocity vector.
If the separation velocity of a fragment detaching from its parent comet in
this direction is $\Delta V_{\rm orb}$, the fragment's orbital velocity,
$P_{\rm frg}$, equals {\it very approximately\/}\footnote{Formula (1)
is derived by differentiating the expression for the velocity in an
elliptic orbit as a function of heliocentric distance and the semimajor
axis of the parent's orbit.  The true effect on the orbital period is
highly nonlinear in $\Delta V_{\rm orb}$.}
\begin{equation}
P_{\rm frg} = P_{\rm par} \!\left( \!1 + \frac{3 \sqrt{2}}{\gamma} \,
	r_{\rm frg}^{-\frac{1}{2}} P_{\rm par}^{\frac{2}{3}} \,
 \Delta \:\!\!V_{\rm orb} + \ldots \right) \!,
\end{equation}
where $\gamma$ is the gravitational constant, $r_{\rm frg}$ is the
heliocentric distance at fragmentation (close to perihelion), and
$P_{\rm par}$ is the orbital period of the parent comet.  When
$P_{\rm par}$ and $P_{\rm frg}$ are in years,{\vspace{-0.06cm}}
$r_{\rm frg}$ in AU, and $\Delta V_{\rm orb}$ in m/s, then
\mbox{$\gamma = 29784.7$ AU$^{\frac{1}{2}}$m/s}.

A fragment does not require momentum exchange to get into a new orbit.
At the instant of its separation from the parent comet whose center of
mass is at a heliocentric distance $r_{\rm frg}$, a fragment's orbital
velocity equals the parent's, but its center of mass is at a slightly
different~helio\-cen\-tric distance \mbox{$r_{\rm frg} \!+\! u_{\rm frg}$}
\mbox{(where $|u_{\rm frg}| \!\ll\! r_{\rm frg})$}.{\vspace{-0.03cm}}
Approximating \mbox{$r_{\rm frg} (r_{\rm frg} \!+\! u_{\rm frg}) \doteq
r_{\rm frg}^2$}, the difference{\vspace{-0.05cm}} in the heliocentric
distance between the two centers of mass is
\begin{equation}
u_{\rm frg} = {\textstyle \frac{1}{2}} r_{\rm frg}^2 \left( \! P_{\rm
	par}^{-\frac{2}{3}} - P_{\rm frg}^{-\frac{2}{3}} \right)
\end{equation}
and the relation between $P_{\rm frg}$ and $P_{\rm par}$ (in years) is
\begin{equation}
P_{\rm frg} = P_{\rm par} \!\left( \!1 - \frac{2u_{\rm frg}}{r_{\rm
 frg}^2} P_{\rm par}^{\frac{2}{3}} \!\right)^{\!\!\!-\frac{3}{2}} \!\!,
\end{equation}
where $u_{\rm frg}$ and $r_{\rm frg}$ are in AU.  A prescribed $P_{\rm
frg}$ requires that \mbox{$u_{\rm frg} \!\sim\! r_{\rm frg}^2$}.  Also,
$u_{\rm frg}$ may be much smaller~than~the distance between the centers
of mass, if their connecting line makes a large angle with the comet's
{\vspace{-0.09cm}}radius vector.  If \mbox{$u_{\rm frg} > \frac{1}{2}
r_{\rm frg}^2 P_{\rm par}^{-\frac{2}{3}}$}, the fragment escapes along
{\vspace{-0.01cm}}a hyperbolic orbit.

It is apparent from Equations (1) and (3),{\vspace{-0.1cm}} respectively,
that in the first approximation \mbox{$P_{\rm frg} \!-\! P_{\rm par}
\sim r_{\rm frg}^{-\frac{1}{2}}$} {\vspace{-0.11cm}}in the momentum
exchange scenario, but \mbox{$P_{\rm frg} \!-\!P_{\rm par} \sim r_{\rm
frg}^{-2}$} {\vspace{-0.05cm}}in the other case, which I refer to below
as a locus scenario.  These correlations are important because the exact
time of fragmentation is not known and in cases of multiple fragmentation
$r_{\rm frg}$ may (because of inertia, for example) change from fragment
to fragment.  More importantly, when a sungrazer's nucleus is assumed to
break up right at perihelion, \mbox{$r_{\rm frg} = q_{\rm par}$}, the
orbital periods of fragments of a Population~I parent (such as X/1106~C1)
and Population~II parent (such as the Chinese comet of 1138) are affected
differently even when their centers of mass are at the same distances
$u_{\rm frg}$ because the perihelion distance in the latter case is
about 1.5~times greater than the perihelion distance in the former
case.\footnote{Here I am replacing the term {\it subgroup\/}, used
by Marsden, with the term {\it population\/}, which I have consistently
been employing in my recent papers on the contact-binary hypothesis of
the Kreutz system (see, e.g., Sekanina 2021, 2022a).}  Specifically,
the orbital periods of Population~I fragments should be{\vspace{-0.03cm}}
stretching $\sqrt{1.5}$~times farther than those of Population~II in case
of the momentum-exchange scenario, but 1.5$^2$~times farther in case
of the locus scenario.  Simulations of the applied sequences of
tidal-fragmentation events may serve to test which of the two proposed
scenarios is more consistent with the data.

\subsection{Simulation of Events of Tidal Fragmentation\\of Second-
and Third-Generation\\Kreutz Sungrazers}
I have employed the two scenarios to simulate events of tidal
fragmentation of some of the largest Kreutz sungrazers that
are the second- and third-generation products of the presumed
progenitor, Aristotle's comet.  Each parent comet has been
assumed to have broken up at perihelion, at distance $q_{\rm
par}$ from the Sun that satisfies the condition \mbox{$r_{\rm frg}
= q_{\rm par}$}.  The orbital period $P_{\rm par}$ is the time
difference between the parent's breakup and its {\it previous\/}
perihelion passage (rounded off to the whole year) and $P_{\rm
frg}$ is the time difference between the fragment's birth and
its {\it following\/} perihelion passage (also rounded off to
the whole year).  In reality $P_{\rm par}$ and $P_{\rm frg}$
are affected by the planetary perturbations, but to a limited
extent only.  The values of $\Delta V_{\rm orb}$ have been
calculated directly from the rigorous expression for the
difference between the two orbital velocities at perihelion,
not from the approximate Equation~(1).

\begin{table*}[ht] 
\vspace{0.2cm}
\hspace{-0.2cm}
\centerline{
\scalebox{0.975}{
\includegraphics{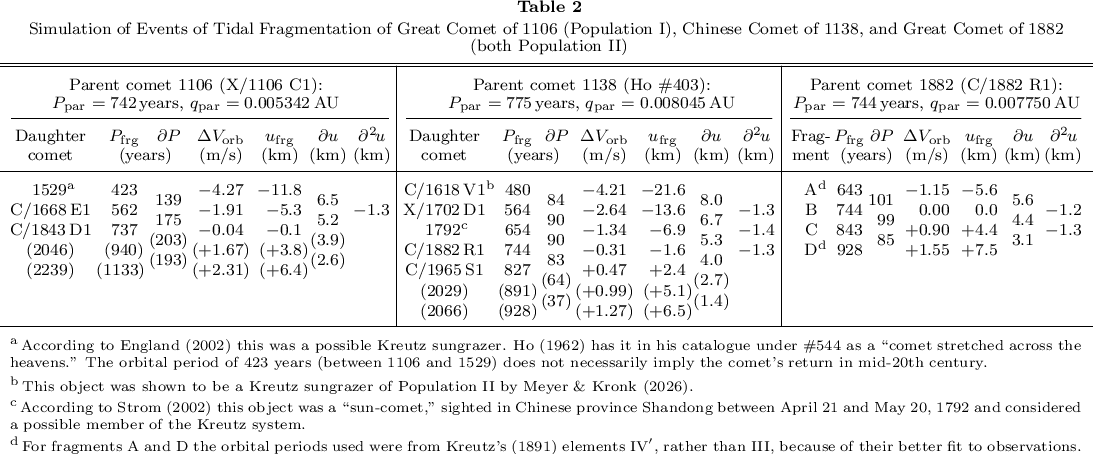}}}
\vspace{0.65cm}
\end{table*}

The results are presented in Table 2, which is divided into three
parts. The first part deals with the simulated fragmentation events
of the Great Comet of 1106, the main second-generation fragment
of Population~I; the second part describes the simulation of
fragmentation products of the Chinese comet of 1138, the main
second-generation fragment of Population~II; and the third part
refers to a simulated scenario of the well-known fragmentation
of the Great September Comet of 1882, the main third-generation
fragment of Population~II.  For each parent the table lists its
true orbital period, $P_{\rm par}$, and the perihelion distance,
$q_{\rm par}$, derived by numerical integration of its motion
(Sekanina \& Kracht 2022).  For each fragment (or predicted
fragment) the table presents, next to its name and true orbital
period, $P_{\rm frg}$:\ the difference between the orbital periods
of the neighboring fragments (or predicted fragments), $\partial
P$; the separation velocity, $\Delta V_{\rm orb}$; and the
difference in the heliocentric distances of the centers of mass
between the fragment and parent, $u_{\rm frg}$, calculated from
Equation~(2).  The last two columns provide, respectively, the
first difference, $\partial u$, and the second difference,
$\partial^2 u$, of the values of $u_{\rm frg}$.  The values of the
second difference are what undoubtedly will catch~the reader's eye.

The simulation of a sequence of tidal-fragmentation events deserves a
commentary because it may have implications for the prognostication of bright
Kreutz sungrazers' appearances in the 21st century and beyond, even though
--- for the reasons already mentioned --- speculation cannot be avoided
in pairing the past objects, with the exception of the comet of 1882 and
Ikeya-Seki (Marsden 1967).  Nonetheless, the table shows that the main
(presumably the most massive) fragment or daughter comet has always its
$u_{\rm frg}$ nearest zero and is close to the middle of the fragment
chain.  The last two rows of the lists of possible fragments for the
parent comets of 1106 and 1138 include parenthesized {\it predictions\/}
of possible future appearances of Kreutz sungrazers, which fit the
established rule based on the essentially fixed value of the second
difference
\begin{equation}
\partial^2 u = -1.3 \pm 0.1 \:\,{\rm km}.
\end{equation}
However remarkable this property of $u_{\rm frg}$ may appear to be,
it is entirely empirical; I see no way to support it by any
argument based on what is known about the Kreutz system, so it is
imperative that the tabulated predictions of future sungrazer
appearances (parenthesized) be assessed with the utmost caution.
Uncertainties in the predicted perihelion times are in any case
{\it at least several years\/} and the possibility that the comet
fails to survive can never be ruled out.

In general, however --- thanks to the low age of the Kreutz system,
an exponential growth in the number of fragments, and a fairly long
lifespan of at least some of them --- occasional appearances of
bright sungrazers in the rest of the 21st century are virtually
guaranteed.  However, the era of truly monumental objects is gone
and will not be back until the time the 19th century giants return
many centuries from now.

Among fragments (or daughter comets) of the Great Comet of 1106,
Table~2 lists, next to the Great March Comet of 1843, the sungrazer
of 1668 (C/1668~E1), which was long suspected to be the previous
return of the 1843 comet.  After collapse of the hypothesis of a
single comet repeatedly returning to perihelion over and over
again (see e.g. Sekanina 2025), it is encouraging to see that
the sungrazers of 1668 and 1843 appear to be likely siblings.
This suggestion is in line with Kreutz's (1901) conclusion that
the approximate positions of the comet of 1668 were fitted by
the orbit of the 1843 sungrazer much better than by the orbit
of the 1882 sungrazer.

I next assume that a comet observed on February~9, 1529, during the
rule of Emperor Jiajing of the Ming Dynasty, was another sibling of
the 1843 sungrazer.  At least England (2002) judges it a possible Kreutz
object.  Ho (1962) has it in his catalogue under No.\,544, described as
a star with a tail (chhang-hsing) ``stretched across the heavens,''
and in Hasegawa's (1980) catalogue its number is 890.  The return of
this possible fragment of the Great Comet of 1106 to perihelion in
1529 demands a short orbital period of 423~years, which of course
could have changed greatly after perihelion.  One thus should not be
surprised that the object did not return around 1952.  In addition,
there always was the danger that, especially if an endpiece in the
fragment chain, the comet was prone to meeting the same fate as
comet Lovejoy.  If it fell apart, it would have returned to
perihelion as a long filament of dwarf sungrazers, starting in the
1950s and continuing into the 21st century and perhaps beyond.  One
thus cannot rule out that many of the dwarf comets detected since
the late 1970s, first by Solwind and SMM and later mostly by SOHO,
have represented the debris of the comet of 1529.  It is well-known
that the majority of the dwarf sungrazers belongs to Population~I, and
a Monte Carlo modeling of the filament led to a conclusion that the
beginning of its existence is likely to have dated back to the 1950s
(Sekanina 2024b).

Because Table 2 shows that the centers of~mass~of~the past fragments
of the Great Comet of 1106 were~\mbox{located} sunward of the parent's
center of mass \mbox{$(u_{\rm frg} \!<\! 0)$}, one should expect the
existence of additional fragments, or daughter comets, to reach
perihelion in the future.  On the assumption of validity of
condition~(4), the earliest object should arrive around 2050, the
following one some time in the 23rd century.

The simulation of tidal fragmentation of the Chinese comet of 1138 is
more robust than the simulation of the comet of 1106, in part because it
includes the widely discussed pair of siblings:\ the 1882 sungrazer and
comet Ikeya-Seki.  But even here the reader can find a weak point, which
is the comet of 1792, one of Strom's (2002) ``sun-comets'' reportedly
sighted between April~21 and May~20 from the Shandong Province of China.
It was the only acceptable candidate that I was able to find between the
Great September Comet of 1882 and comet X/1702~D1, another probable
member of Population~II according to Kreutz (1901).

Meyer \& Kronk (2026) recently concluded that the orbital elements
for comet C/1618~V1, published by Landgraf (1985) were incorrect
and that records on this object's observations by Jesuits in India
confirmed that it was a Kreutz sungrazer.  It appears that this
comet could be yet another fragment of the Chinese comet of 1138.
A similar argument on the parent's center of mass as in the case
of the 1106 comet supports a suggestion that further sizable
fragments should be on their way to perihelion.  A striking
feature of this prediction is that the earliest of them is
expected as early as a few years from now, around 2030 (again
with implied uncertainty of at least several years), the next
one some four decades later.

The prediction of an impending appearance of a potentially bright
sungrazer merits two comments.  One~is~that recently I did already
call attention to a forthcoming event of this kind (Sekanina 2025).
Independent of the algorithm dictated by condition~(4), my
prediction was~the year 2027, agreeing with the current prognosis
much better than is the uncertainty.  The other comment concerns
the physical state in which the predicted object might arrive;
catastrophic disruption shortly before perihelion or in its proximity
is entirely unpredictable.  In this particular case a difficult
question arises whether the dwarf comet C/2024~S1, which completely
sublimated away before perihelion and whose appearance I discussed
elsewhere (Sekanina 2024c), was a precursor to the predicted object
or the decrepit object itself.

Effects of tidal fragmentation of the Great Comet of 1882 were studied
extensively by Kreutz (1888, 1891) and on a number of occasions more
recently by others.  Here I comment only on the data in Table~2.  For
fragments~B (the main one) and C the orbital periods were taken from
Kreutz's (1891) elements~III, for fragments A and D from elements
IV$^\prime$, which appeared to fit Kreutz's normal places better.
The osculating values were converted to the barycentric values using
the data on the comet's semimajor axis published in Marsden \&
Williams' (2008) catalogue, allowing for the differences (i)~between
the osculation epochs adopted by Kreutz on the one hand and by Marsden
\& Williams on the other hand; and (ii)~between Kreutz's (1891)
nonrelativistic elements and Hufnagel's (1919) relativistic elements
for nucleus~B, which were adopted in the catalogue and whose accuracy
I have been questioning.

\begin{table*}[ht] 
\vspace{0.2cm}
\hspace{-0.25cm}
\centerline{
\scalebox{1}{
\includegraphics{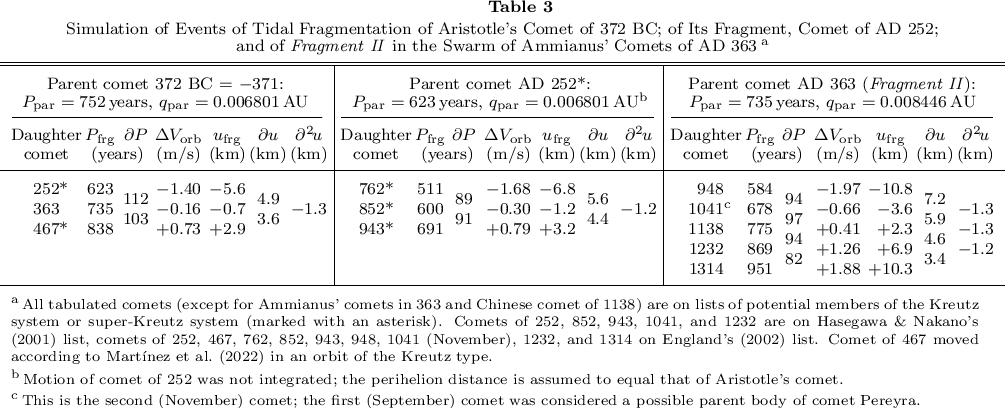}}}
\vspace{0.78cm}
\end{table*}

Even a rather perfunctory glance at Table 2 suffices to show that
the differences $\partial P$ between the orbital periods of
neighboring daughter comets of Population~I (the 1106 parent)
are on the average at least twice as long (${\partial P}$ up to
203~years) as those of Population~II (the 1138 and 1882 parents;
$\partial P$ as much as 101~years).  This suggests that the
temporal relations among the tabulated tidal-fragmentation objects
are consistent with the proposed locus scenario rather than the
momentum-exchange scenario.  The net outcome in practice is a
higher incidence rate of bright sungrazers that belong to
Population~II in comparison to Population~I.  This expectation
contrasts with the situation among the dwarf sungrazers, where
the Population~I objects dominate.

\subsection{Simulation of Events of Tidal Fragmentation of\\Aristotle's
Comet and Its First-Generation Products}
All comets discussed in Section 5.1 and listed in Table~2 were
genuine Kreutz sungrazers, fragments of one of the two lobes of
Aristotle's comet after their separation, as postulated by the
contact-binary hypothesis~(\mbox{Sekanina} 2021).  By contrast, the
simulation of tidal-fragmentation events of Aristotle's comet itself
and its first-generation products is more complicated because these
episodes involved not only the genuine Kreutz sungrazers but also
objects that had separated from the progenitor before the birth of the
Kreutz system.  Indeed, once Aristotle's comet was gradually perturbed
into its sungrazing orbit by a mechanism described by\vspace{-0.03cm}
Bailey et al.\ (1992), it unquestionably became subject to tidal
fragmentation.\footnote{It should be emphasized that the high
probability of fragmentation of Aristotle's comet has nothing in
common with the claim of Ephorus of Cyme that in 372~BC he {\it
saw\/} the comet to split in two.  This report, often mentioned
even in scientific papers, can safely be dismissed, because there
is absolutely no way to see a genuine fragmentation event with the
naked eye.  At best one can observe a sudden brightening (outburst
or flare-up) of the comet that {\it may\/} accompany the breakup.}
At least some of the fragments were massive enough to survive one
or more revolutions about the Sun.  Moreover, at least some of the
larger among these fragments {\it continued to tidally break up\/}
at perihelion during their following returns.  Given the enormous
range of orbital periods of such fragments of the first and second
generations born during the returns of Aristotle's comet in the
4th, 12th, and even 20th century BC, their total number in the
first millennium AD must have been fairly large to say the least.
None of these sungrazers was a member of the Kreutz system
(because they did not originate from one of the lobes {\it
after\/} their separation), yet they all did move in orbits
typical for the Kreutz comets.

In a paper in which I called attention to these objects, I
referred to them as the {\it super-Kreutz system\/} (Sekanina
2023).  I pointed out that, in the context of the contact-binary
hypothesis, virtually all comets before the beginning of the
9th century moving in Kreutz-like orbits were members of the
super-Kreutz system.  On their lists of suspects, Hasegawa \&
Nakano (2001) provide nine such objects, England (2002) as many
as 22.  The only exception was an apparent swarm of spectacular
daylight comets in AD~363, recorded by Ammianus Marcellinus.
This was the first arrival to perihelion of the two separated
and fragmented lobes (Sekanina 2022b), offering the earliest
opportunity to their tidal breakup.

Starting with the 9th or 10th century, the Kreutz and super-Kreutz
sungrazers got all mixed together.  However, it is likely that the
Kreutz comets soon had the upper hand, because they contained most
of the mass of Aristotle's comet.  To have a look at possible
examples of the super-Kreutz sungrazers and their co-existence
with the Kreutz sungrazers, simulated cases of tidal fragmentation
of Aristotle's comet are shown on the left of Table~3, which has
format of Table~2.  I assume that before the comet gave birth to
the Kreutz system, at least two major fragments separated from
it at perihelion in 372~BC:\ one may have returned to perihelion
in AD~252, the other in AD~467.  The first is on the list of possible
sungrazers by Hasegawa \& Nakano (2001), the other was considered
a likely sungrazer by Mart\'{\i}nez et al.\ (2022).  Both were
thought to be possible sungrazers by England (2002).  If this
hypothesis is correct, the return of Aristotle's comet lasted over
two centuries and the fragments at either end were super-Kreutz
sungrazers, identified in Table~3 with an asterisk.  Between them
was the first arrival of the Kreutz system at perihelion.

The middle part of Table 3 contains a simulated case of tidal
fragmentation of the super-Kreutz sungrazer of AD~252 into three
pieces\footnote{As the motion of this comet from 372~BC to AD~252
was not integrated, its perihelion distance could not be determined;
it was assumed to equal the perihelion distance of Aristotle's
comet.} (each on a list of suspects), reaching perihelion in
762, 852, and 943, respectively.  Least information is available
on the Byzantine comet of 762, which according to Pingr\'e (1743)
was described by Theophanes, a Constantinople historian, to have
``form of a beam.''  The comets of 852 and 943 are well documented
by Hasegawa \& Nakano (2001), the former observed in Japan and
China, with the perihelion time estimated on March~10\,$\pm$\,2~days,
the latter reported from China and Germany, at perihelion on
October~27\,$\pm$\,1~day.

It is expected that a number of tidal fragmentation events took
place when the swarm of first-generation fragments of the
separated lobes of Aristotle's comet reached perihelion in 363.
While the actual number of sizable pieces that were arriving is
unknown, reconstruction of the Kreutz system architecture from
the orbital properties of hundreds of SOHO dwarf sungrazers led
to a conclusion that Ammianus' swarm of daylight comets may have
consisted of up to nine or ten massive objects passing through
perihelion over a period of several days (Sekanina 2022a, 2022b).
If most of them fragmented, each into, say, \mbox{3--5}~major
parts, the number of sizable objects (a few kilometers or more
across) leaving the perihelion in 363 could easily have run into
dozens.

Of particular interest was the fragmentation of the two largest
masses, each believed to have carried much of what used to be
Lobe~I and Lobe~II, respectively; I have been referring to these
as \mbox{\it Fragment~I\/} and \mbox{\it Fragment II\/}.\footnote{See
the early chart of the pyramidal architecture of the Kreutz system
in Figure~3 of Sekanina (2022b) and its updated version in Figure~3
of Sekanina (2025).}  The perihelion arc of the orbit in AD~363
was the~birth place of the Great Comet of 1106 (X/1106~C1),~the
largest surviving piece of {\it Fragment~I\/}, and likewise the
birth place of the Chinese comet of 1138, the largest surviving
piece of \mbox{\it Fragment~II\/}.  Tidal fragmentation of {\it
Fragment II\/} is simulated on the right of Table~3.

Not much is known about the Japanese comet of 948, which was found
on March~2 in the southwest, and on the European comet of 1314,
which was seen in October (Pingr\'e 1743; England 2002).  On the
other hand, of much interest is the story with a comet, or comets, of
1041.  Hasegawa \& Nakano (2001) have said that this Chinese-Korean
comet was under observation over a period longer than three months
and in their list of 24 Kreutz candidates it is by far the
intrinsically brightest object.  It is more likely that these
were two comets, as presented by England (2002):\ the first,
Korean-Byzantine, discovered in September in the east, with a
45$^\circ$\,long tail and visible for more than 20~days;~and the
second, Korean, discovered between late October and late November
in the east, with a tail of similar length and seen for more
than 10~days.  They are~also~listed~as separate objects by Ho (1962)
under Nos.\ 371 and 372 and Hasegawa (1980) under Nos.\ 577
and 578.  \mbox{Sekanina} \&~Kracht (2022) speculated that the first
one might have been Pereyra's parent comet; the second is suggested
here to be a sibling of the Chinese comet of 1138.  The
Chinese-Japanese comet of 1232 has likewise been well documented,
seen in the east starting with October~17 and with a tail length
exceeding 60$^\circ$ in November. According to Hasegawa \& Nakano
(2001) the comet reached perihelion on October 14\,$\pm$\,2 days.

Table 3 shows that the condition (4) applies, at least to the degree
demonstrated by the limited number of selected entries, to fragments
of Aristotle's comet,~a~super-Kreutz sungrazer, and a daylight comet
of Ammianus.  Yet, the most robust case remains to be that of
the~frag\-ment nuclei of the Great September Comet of 1882,~fur\-ther
supported by the pairs of C/1882~R1 vs C/1965~S1, C/1668~E1 vs C/1843~D1,
and C/1618~V1 vs X/1702~D1, which appear to strengthen the validity of
condition~(4) and the locus scenario.

\section{Final Comments and Conclusions}
A broad picture of the Kreutz system over past two millennia suggests
a nearly eight-century long cycle that has an apparent tendency
to repeat itself.  \mbox{Particularly} impressive sungrazers are believed
to have reached peri\-helion in the second half of the 4th century,
in~the~first half of the 12th century, and in the course
of~the~19th~century.  In the context of the recently proposed
contact-binary hypothesis (Sekanina 2019), this information is
compatible with more comprehensive evidence on bright sungrazers
observed over the past 200~years.  That evidence is interpreted
to mean that the appearances of {\it two\/} spectacular objects
--- the Great March Comet of 1843 and the Great September Comet
of 1882 --- less than 40~years apart was not a coincidence:\
their temporal separation was an accumulated effect over two
revolutions about the Sun of a difference between their
initial orbital periods.  Neither was it a coincidence that the
20th~century~witnessed intrinsically~\mbox{bright --- but not
spectacular}~\mbox{---}~sungraz\-ers,~whereas~the 21st
century~has~so~far~been~a~\mbox{disaster}.
These~\mbox{developments}~are~fully~\mbox{consistent} 
with~a~\mbox{long-term} down\-hill~trend~in~the~spatial
distribution~of~bright~Kreutz comets starting in the late 1800s.

Overlapping this trend are clusters of bright sungrazers, whose
existence was called attention to by Marsden in 1967.  They are
poorly defined, but appear to be divided into three kinds:\ genuine,
mixed, and fortuitous; the last ones seem to dominate.

Dwarf Kreutz sungrazers, detected primarily by the coronagraphs on
board the SOHO spacecraft, also tend to cluster, and I refer to
their assemblages as clumps.  Unlike the clusters of bright comets,
the clumps are overwhelmingly genuine in nature.  Triggered by
cascading fragmentation of initially larger objects into ever
smaller ones along the entire orbit, the dwarf sungrazers are
mostly boulders meters to tens of meters across, all of which
get destroyed by sublimation shortly before reaching perihelion,
thus marking the terminal phase of evolution of the Kreutz system.

The extremely complex architecture of the Kreutz system is a
product of two kinds of fragmentation.  Dominant throughout
nearly the entire orbit is essentially spontaneous breaking
up of sungrazers' nuclei at very low separation velocities,
possibly rotational in nature.  The greater the heliocentric
distance of the fragmentation event is, the more strongly
affected are the fragments' angular elements and/or perihelion
distances (depending on the separation velocity vector).

Only at close proximity to perihelion, within a few solar 
radii from the photosphere, are the Kreutz comets subjected
to tidal fragmentation, if their nuclei are sizable enough
and their cohesion poor.  Unlike in cases of nontidal
fragmentation, no momentum exchange is necessary for
fragments to end up in orbits of considerably different
periods (but essentially the same angular elements and
perihelion distance) because at the instant of separation
the centers of mass of the parent comet and the fragments
are at slightly different heliocentric distances, yet their
orbital velocities are equal.  For a fragment to be perturbed
into an orbit of a prescribed period it is required that the
deviation of the fragment's center of mass from the parent's
center of mass along the radius vector, $u_{\rm frg}$, vary
as a square of the heliocentric distance at fragmentation.

It turns out that this simple mechanism is extremely effective in
distributing fragments into orbits of widely ranging periods, which
makes tracking fragments of different generations increasingly
difficult, as the number of their revolutions about the Sun
increases.  Simulated in terms of $u_{\rm frg}$ has been tidal
fragmentation of six bright sungrazers, including the Great Comet
of 1882 (whose fragmentation products were extensively observed),
the Great Comet of 1106, and Aristotle's comet.  Also included
has been one so-called super-Kreutz comet, which was not a member
of the Kreutz system proper.  A remarkable property of all six
simulations was that the second difference of $u_{\rm frg}$,
derived from the orbital periods of ``neighboring'' fragments
on the assumption that the parent broke up exactly at
perihelion, always equaled \mbox{$\partial^2 u = -1.3 \pm
0.1$ km}.

The significance of this result is exceptional because of its
potential ramifications for the prognostication of appearances of
bright Kreutz sungrazers for the rest of the 21st century and
beyond.  Arrival of one of them, of Population~II, is predicted
to be impending.  However, because the applied procedure is
empirical and not supported by a theoretical argument, the
listed predictions of future sungrazer appearances, including
the forthcoming one, should be judged with the utmost caution.

\vspace*{-0.1cm}
\begin{center}
{\footnotesize REFERENCES}
\end{center}
\vspace*{-0.2cm}
%
\parbox{8.63cm}{\footnotesize
Bailey,\,M.\,E.,\,Chambers,\,J.\,E.,\,\&\,Hahn,\,G.\,1992,\,Astron.\,Astrophys.,
 {\hspace*{0.25cm}}257, 315 \\[0.03cm]
England, K.\ J.\ 2002, J.\ Brit.\ Astron.\ Assoc., 112, 13 \\[0.03cm]
Fabritius, W.\ 1883, Astron.\ Nachr., 105, 287 \\[0.03cm]
Frisby, E.\ 1883, Astron.\ Nachr., 104, 159 (erratum:\ 283); also:\ Na-
 {\hspace*{0.25cm}}ture, 27, 226 \\[0.03cm]
Hasegawa, I.\ 1980, Vistas Astron., 24, 59 \\[0.03cm]
Hasegawa, I., \& Nakano, S.\ 2001, Publ.\ Astron.\ Soc.\
 Japan,~53,~931 \\[0.03cm] 
Ho, P.-Y.\ 1962, Vistas Astron., 5, 127 \\[0.03cm]
Holetschek, J.\ 1892, Astron.\ Nachr., 129, 323 \\[0.03cm]
%
%
%
%
%
%
Hufnagel, L.\ 1919, Astron.\ Nachr., 209, 17 \\[0.03cm]
Kreutz, H.\ 1888, Publ.\ Sternw.\ Kiel, No.\ 3 \\[0.03cm]
Kreutz, H.\ 1891, Publ.\ Sternw.\ Kiel, No.\ 6 \\[0.03cm]
%
%
Kreutz, H.\ 1901, Astron.\ Abhandl., 1, 1 \\[0.03cm]
Landgraf, W.\ 1985, Sterne, 61, 351 \\[0.03cm]
%
%
Marsden, B.\ G.\ 1967, Astron.\ J., 72, 1170 \\[0.03cm]
Marsden, B.\ G.\ 1989, Astron.\ J., 98, 2306 \\[0.03cm]
Marsden, B.\ G., \& Williams, G.\ V.\ 2008, Catalogue of Cometary
 {\hspace*{0.25cm}}Orbits 2008 (17th ed.).  Cambridge, MA:\ IAU Central
 Bureau~for
 {\hspace*{0.25cm}}Astronomical Telegrams and Minor Planet Center,
 195pp\\[0.03cm]
Marsden,\,B.\,G., Sekanina,\,Z., \& Everhart,\,E. 1978,
 Astron.\,J.,~83,~64 \\[0.03cm]
%
%
Mart\'{\i}nez, M.\ J., Marco, F.\ J., Sicoli, P., \& Gorelli, R.\ 2022,~Icarus,
 {\hspace*{0.25cm}}384, 115112 \\[0.03cm]
Meyer, M., \& Kronk, G.\ W.\ 2026, J.\ Astron.\ Hist.\ Herit., 29, 93 \\[0.03cm]
Morrison, J.\ 1883, Mon.\ Not.\ Roy.\ Astron.\ Soc., 44, 49 \\[0.03cm]
Pingr\'e,\ A.\ G.\ 1783, Com\'etographie ou Trait\'e Historique et Th\'eo-
 {\hspace*{0.25cm}}rique des Com\`etes. Paris:\ L'Imprimerie Royale,
 630pp \\[0.03cm]
Sekanina, Z.\ 2000, Astrophys.\ J., 542, L147 \\[0.03cm]
Sekanina, Z.\ 2021, eprint arXiv:2109.01297 \\[0.03cm]
Sekanina, Z.\ 2022a, eprint arXiv:2212.11919 \\[0.03cm]
Sekanina, Z.\ 2022b, eprint arXiv:2202.01164 \\[0.03cm]
Sekanina, Z.\ 2023, eprint arXiv:2305.08792 \\[0.03cm]
Sekanina, Z.\ 2024a, eprint arXiv:2401.00845 \\[0.03cm]
Sekanina, Z.\ 2024b, eprint arXiv:2404.00887 \\[0.03cm]
Sekanina, Z.\ 2024c, eprint arXiv:2411.12941 \\[0.03cm]
Sekanina, Z.\ 2025, eprint arXiv:2503.15467 \\[0.03cm]
Sekanina, Z., \& Chodas, P. W. 2007, Astrophys.\ J., 663, 657 \\[0.03cm]
%
%
%
Sekanina, Z., \& Kracht, R.\ 2022, eprint arXiv:2206.10827 \\[0.03cm]
Strom, R.\ 2002, Astron.\ Astrophys., 387, L17}
\vspace{-0.09cm}
\end{document}